\documentclass[fleqn,usenatbib]{mnras}

\usepackage{newtxtext,newtxmath}
\usepackage[T1]{fontenc}

\DeclareRobustCommand{\VAN}[3]{#2}
\let\VANthebibliography\thebibliography
\def\thebibliography{\DeclareRobustCommand{\VAN}[3]{##3}\VANthebibliography}

\usepackage{graphicx}	% Including figure files
\usepackage{amsmath}	% Advanced maths commands
\usepackage{xcolor}
\usepackage{multirow}
\usepackage{subcaption}
\usepackage{mwe}

\title[Constraining gas metal mixing strength]{Constraining gas metal mixing strength in simulations using observations of the Milky Way's disc}

\author[J. Sarrato Alós et al.]{
J. Sarrato-Alós,$^{1,2}$\thanks{E-mail: jsarrato@iac.es (JSA)}
C. Brook,$^{1,2}$
and A. Di Cintio$^{1,2}$
\\
$^{1}$Instituto de Astrofísica de Canarias, Calle Via Láctea s/n, E-38205 La Laguna, Tenerife, Spain\\
$^{2}$Universidad de La Laguna, Avda. Astrofísico Fco. Sánchez s/n, E-38206 La Laguna, Tenerife, Spain\\
}

\date{Accepted XXX. Received YYY; in original form ZZZ}

\pubyear{2022}

\begin{document}

\newpage
\label{firstpage}
\pagerange{\pageref{firstpage}--\pageref{lastpage}}
\maketitle

\begin{abstract}
This work explores the mixing rate of metals in the interstellar medium (ISM), comparing observational constraints from our solar neighbourhood to high resolution cosmological hydrodynamical simulations of Milky Way (MW)-like galaxies. The mixing rate, described by the coefficient C, is varied in simulations between 0 and 0.05, with resultant simulated galaxies compared to observations of metallicity dispersion in
young star clusters, HII regions and neutral gas in the disc of the MW. A value of C between 0.003125 and 0.0125 is found to self-consistently match a range of observables, with a best estimate of C=0.0064$\pm$0.0004. We demonstrate that the relationship between metal dispersion in young stars, HII regions and neutral gas, versus the coefficient C, can be described by a power law.
These constrained mixing rates infer a comparatively well mixed ISM in the solar neighbourhood, at odds with some recent observations that have reported a highly inhomogeneous ISM.
The degree of mixing suggested by this work is lower than what often employed in many hydrodynamical simulations.
Our results have implications for studying the metallicity distribution of stars as well as of gas in the interstellar and circumgalactic media.

% Recent observations of the metal dispersion found in stars and gas around our solar neighbourhood are compared, for the first time, with a set of cosmological hydrodynamical simulations of the Milky Way (MW), covering a wide range of metal mixing strength. We find that none of the explored simulation is able to reproduce the recently reported high metallicity dispersion in the solar neighbourhood %with other observations. , however, we show how all other sources of observational data are consistent with each other via comparison with simulations of similar mixing strength \ary{i dont understand this. what other observations?be specific. Hi?H2? stars?gas? where?}.% Our comparisons may be influenced by processes other than metal mixing which affect the resulting metallicity dispersion, e.g. supernovae metal returns and differences in star formation history. .this is more for a discussionin the paper, not for abstract

% We find that a MW simulation, using GASOLINE's implemented sub-grid metal mixing model with a relatively small %scaling constant metal mixing coefficient of C $\sim$ 0.006 (as opposite  to the more commonly used value of C = 0.05), provides the necessary metal mixing strength to reproduce nicely the observations studied in this work. This demonstrates that local observations of both  stars and  gas metal dispersion can be used jointly with high resolution  simulations of our Galaxy to constrain the level of metal mixing expected.

\end{abstract}


%%%%%%%%%%%%%%%%%%%%%%%%%%%%%%%%%%%%%%%%%%%%%%%%%%

%%%%%%%%%%%%%%%%% BODY OF PAPER %%%%%%%%%%%%%%%%%%

\section{Introduction}
Spatial variations of metallicity in the interstellar medium (ISM) within the disc of the Milky Way (MW) can provide fundamental constraints on the importance of gas mixing processes and their rates. Such constraints can be applied to the rates of metal mixing that are included within hydrodynamical simulations of galaxy formation. 

However, the degree of metallicity ([M/H]) variations in the Galactic ISM has been a matter of debate, providing a complication for using such observations as constraints on simulations. Whilst gas in the solar neighbourhood is generally assumed to be well mixed and to have metallicities similar to that of the Sun, \cite{decia} found significant variability (up to 1 dex) in the [M/H] values of the neutral ISM surrounding the Sun, including metallicities as low as [M/H] = -0.78. 

This finding was challenged by \cite{Esteban_2022}, who measured the spread in metallicity using various tracers of the ISM including Cepheids, young star clusters and HII regions, and concluded that the chemical composition of the ISM in the solar neighbourhood is relatively homogeneous. These results were more in line with previous findings in the field, which have indicated a well mixed ISM once radial variance is accounted for, using HII regions \cite[e.g.][]{Esteban18,ArellanoCordova} or young stars and star clusters \cite[e.g.][]{SimonDiaz10,Nieva12,Luck2018,Donor_2020}. 

Recently, \cite{Ritchey} performed a similar analysis to that of \cite{decia} and determined that the metallicity of the ISM was relatively homogeneous, providing further support to the classical vision of a well mixed solar neighbourhood. 

On the theory side, hydrodynamical cosmological simulations provide self-consistent depictions of the formation and evolution of galaxies, allowing the study of the distribution of chemical elements in stars and gas. Metals influence the evolution of the simulated galaxies, including controlling gas cooling and therefore affecting star formation. Along with metal returns via supernovae and stellar winds, metal mixing and turbulence determine the distribution of different chemical elements throughout the simulation.

In particular, Lagrangian codes, such as Smooth Particle Hydrodynamics (SPH), do not mix entropy, heat, and other tracers such as metals between particles, and for this reason they yield unphysical results \cite[e.g.][]{wadsley}. Such codes need to include metal mixing via an explicit addition of diffusion using sub-grid models. Mixing rates in Lagrangian codes have often been calibrated to reproduce rates found in Eulerian codes, which naturally include metal mixing \cite[e.g.][]{wadsley}. 

Several studies have attempted to constrain the mixing rates in hydrodynamical simulations using Local Group observations. For example, \cite{alfascattering} used idealised simulations to highlight the need to include metal mixing in Lagrangian codes for reproducing the scatter of $\alpha$-elements in the MW and local dwarf galaxies, whilst \cite{escala} show that their cosmological hydrodynamical simulations best reproduce the width of the metallicity distribution function (MDF) in Local Group dwarf galaxies when adding sub-grid metal diffusion.

\cite{rennehan19} implemented a dynamic calculation of the mixing rate based on local properties of the simulation, and compared this method to the classical choice of a constant metal mixing strength, showcasing significant differences in the MDFs of gas phases. \cite{Rennehan_2021} also demonstrated the impact of changing metal mixing on the overall outcomes of hydrodynamical simulations by performing a comparative analysis between the Smagorinsky model \citep{smagorinsky_1963} and the gradient model \citep{gradient,Hu_2020} for metal diffusion in hydrodynamical simulations of galaxies. While the Smagorinsky model assumes isotropic diffusion and represents unresolved sub-grid turbulence through shear fluctuations, the gradient model avoids such assumptions by approximating sub-grid turbulence fields using Taylor expansion. They also tested the aforementioned dynamic mixing rate against several values for fixed mixing rates, finding that using constant mixing strengths is a well-behaved option for avoiding the extra computation needed in dynamic models.

For this letter we make a suite of simulations of a MW analogue galaxy, with varying mixing rates. Our goal is to constrain the necessary metal diffusion strength in SPH simulations using observational data from the MW. For this purpose, we use 8 runs of a simulated galaxy whose initial conditions come from the NIHAO project \citep{Wang_2015}, each with different metal diffusion strength. We compare data from the suite of simulated galaxies to observations of various sources: \cite{Donor_2020}, \cite{ArellanoCordova}, \cite{decia} and \cite{Ritchey}. The comparisons include chemical abundances of stars and gas in the solar neighbourhood and in the MW's disc. We aim to determine which of the simulated galaxies better reproduces observed data, and to use this test to obtain constraints on the necessary metal diffusion strength in simulations. 

Since \cite{decia}'s and \cite{Ritchey}'s observations are incompatible, we also aim to use the comparisons with simulations to provide an insight on the discrepancy.

We describe the main properties of simulated Milky Way galaxies in section \ref{sect:sim}, including detailed explanation of how metal diffusion is implemented. The observational and simulated sample of stars and gas used for this analysis are presented in section \ref{sect:obs} and \ref{sec:sample}, respectively. In section \ref{sec:results} we present our results, and we discuss them and summarise our conclusions in section \ref{sec:disc}.
%The discrepancy in metal dispersion in the solar neighbourhood among these works is one of the topics addressed in this paper via the use of cosmological hydrodynamic simulations. Cosmological simulations constitute an important tool for the prediction and study of physical phenomena in the Universe. They are specially useful for unveiling the mechanisms of formation and evolution of complex astronomical structures such as galaxies and galaxy clusters. In this work, we use them to study the distribution of chemical elements in the vicinity of the Sun. 
%\cite{wadsley} were able to solve the physical problems they found in SPH simulations by adding a diffusion rate to SPH simulations. This allows for mixing of scalar quantities (temperature, entropy, chemical abundances...) in gas. Shortly after, \cite{shen} proposed a slightly different diffusion model that is meant to better account for turbulence. The latter has been implemented into the newest version of the well known GASOLINE code (\cite{gasoline2}), which is a solver for cosmological hydrodynamic simulations. The mixing strength of both models is tuneable through a constant C that we aim to calibrate in this work. The higher C is, the more efficient metal diffusion is. More details on the mixing models in section \ref{sec:diffmodels} and in the original papers.
\section{Simulations}
\label{sect:sim}
We re-simulate a MW analogue, g8.26e11 from the NIHAO project \citep{Wang_2015}, with the SPH code \textit{ESF-GASOLINE2} \citep{Wadsley_2017}, varying the coefficient that controls the mixing rate. The simulated galaxy has total mass within the range attributed to the MW \citep[e.g.][]{fritz2020}, with a stellar mass that is at the low end of the range estimated by \cite{Licquia_2015}. It also went through its last significant merger at redshift z $\sim$ 1.5, which resembles the Gaia-Enceladus \citep{Brook_2003,Gaia_Enceladus,helmi18} merger event in our Galaxy. This galaxy is a good MW analogue in terms of kinematics, showing  good agreement with the observed rotation curve of our galaxy \citep{buckMWlike}. It structurally resembles the Galaxy, presenting a spiral morphology including not only thin and a thick disc \citep{MW_struct}, but also a classical and a pseudo bulge, and a stellar halo, with properties of all components within the expected observational ranges for shapes, velocities, rotational support and specific angular momenta \citep{mwkinematics}. The chemical distribution of stars in the galaxy shows MW-like separation between the thin and thick disc, as shown in \cite{Buck_2021}, where versions of this simulation with a different implementation of chemical evolution are compared to GALAH data \citep{galah}. One notable disagreement with respect to observations is the scatter of chemical abundances, which is addressed in this paper. 

The code includes ultraviolet heating, ionisation and metal cooling \citep{shen}. Star formation and feedback follow the model used in the Making Galaxies In a Cosmological Context simulations (MaGICC, \citealt{stinson13}), reproducing galaxy scaling relations over a wide mass range \citep{brook12b}. The density threshold for star formation is $n_{\rm th}$$>$$10.3 \rm cm^{-3}$ and a \citet{Chabrier03} IMF is used. Energy from stars is injected into the ISM, modelled using blast-wave supernova feedback \citep{stinson06} and early stellar feedback from massive stars. Particle masses and force softenings allow the mass profile to be resolved to below 1$\%$ of the virial radius, which ensures that galaxy half-light radii are well resolved.

The cosmology adopted for NIHAO is maintained, coming from \cite{planckcosmo}: H$_{\text{0}}$ = 100h km s$^{\text{-1}}$ Mpc$^{\text{-1}}$ with h = 0.671, $\Omega_{\text{m}}$ = 0.3175, $ \Omega_{\Lambda}$ = 0.6824, $\Omega_{\text{b}}$ = 0.049 and $\sigma_{\text{8}}$ = 0.8344.

% \usepackage{graphicx}
\begin{table}
\caption{Properties of simulated Milky Way galaxies. The original NIHAO one is indicated in the bottom line. The table shows the diffusion coefficient C, the halo mass M$_{200}$, the gas and stellar mass, and the virial radius R$_{200}$.}
\label{tab:galproperties}
\resizebox{\columnwidth}{!}{%
\begin{tabular}{cccccc}
\hline
C        & \begin{tabular}[c]{@{}c@{}}M$_{200}$\\ (10$^{12}$ M$_{\odot}$)\end{tabular} & \begin{tabular}[c]{@{}c@{}}Gas mass\\ (10$^{11}$ M$_{\odot}$)\end{tabular} & \begin{tabular}[c]{@{}c@{}}Star mass\\ (10$^{10}$ M$_{\odot}$)\end{tabular} & \begin{tabular}[c]{@{}c@{}}R$_{200}$\\ (kpc)\end{tabular}  \\ \hline
0        & 1.27                                                                        & 1.16                                                                       & 4.58                                                                        & 337                                                                                                          \\
0.0002   & 1.27                                                                        & 1.16                                                                       & 4.20                                                                        & 337                                                                                                       \\
 0.001    & 1.28                                                                        & 1.18                                                                       & 4.57                                                                        & 338                                                                                                          \\
 0.003125 & 1.27                                                                        & 1.08                                                                       & 4.92                                                                        & 337                                                                                                         \\
 0.00625  & 1.27                                                                        & 1.10                                                                       & 5.05                                                                        & 337                                                                                                          \\
 0.0125   & 1.27                                                                        & 1.07                                                                       & 4.62                                                                        & 337                                                                                                       \\
 0.025    & 1.26                                                                        & 1.01                                                                       & 4.71                                                                        & 336                                                                                                         \\ \hline
0.05     & 1.25                                                                        & 1.00                                                                       & 4.74                                                                        & 335                                                                                                         \\ \hline
\end{tabular}%
}
\end{table}

\subsection{Sub-grid diffusion models}
\label{sec:diffmodels}
The sub-grid metal mixing model implemented in ESF-GASOLINE2 was developed in \cite{shen}. In order to implement diffusion into SPH simulations, \cite{wadsley} and \cite{shen} proposed sub-grid models based on \cite{smagorinsky_1963}'s study for the terrestrial atmosphere.

\cite{wadsley}'s model uses a diffusion rate D (equation \ref{diffrate}) which depends on the pairwise velocity of gas particles $\Delta\textrm{v}$ and the smoothing length $\textrm{h}_{\textrm{SPH}}$, with a diffusion coefficient C that controls the variation of a scalar quantity A over time due to diffusion, as described in equation \ref{diffderivative}.
\begin{equation}
    \label{diffrate}
    \textrm{D} = \textrm{C} \Delta\textrm{v} \textrm{h}_{\textrm{SPH}}
\end{equation}
\begin{equation}
    \label{diffderivative}
    \left.\frac{\textrm{d}A}{\textrm{d}t}\right\vert_{Diff} = \nabla \left( D \nabla A\right)
\end{equation}
\cite{shen} proposed a slightly different turbulent mixing model, where the diffusion rate is calculated according to equation \ref{newdiffrate}. In this model, the so-called trace-free shear tensor $S_{ij}$ is used to calculate the diffusion rate, rather than using pairwise velocities. This change is implemented in order to better reproduce mixing due to turbulent gas flows.
\begin{equation}
    \label{newdiffrate}
    \textrm{D} = \textrm{C} \mid S_{ij} \mid \textrm{h}^{2}_{\textrm{SPH}}
\end{equation}
The rate at which metals are diffused in simulations is a key factor affecting the metal distribution of gas and stars in the galaxy. For this reason, a physically motivated tuning of metal diffusion strength is a need in SPH simulations. 

Initially, the diffusion strength of the models implemented into SPH simulations was set to reproduce the results of Eulerian simulations. \cite{wadsley} and \cite{shen} used this approach, using the constraint provided by the entropy profile of Eulerian simulations of galaxy clusters, and concluded that values of C of $\sim$0.1 and $\sim$0.05 are sufficient in their respective diffusion rate equations. 

\begin{figure*}
    \centering
    \includegraphics[scale=.2]{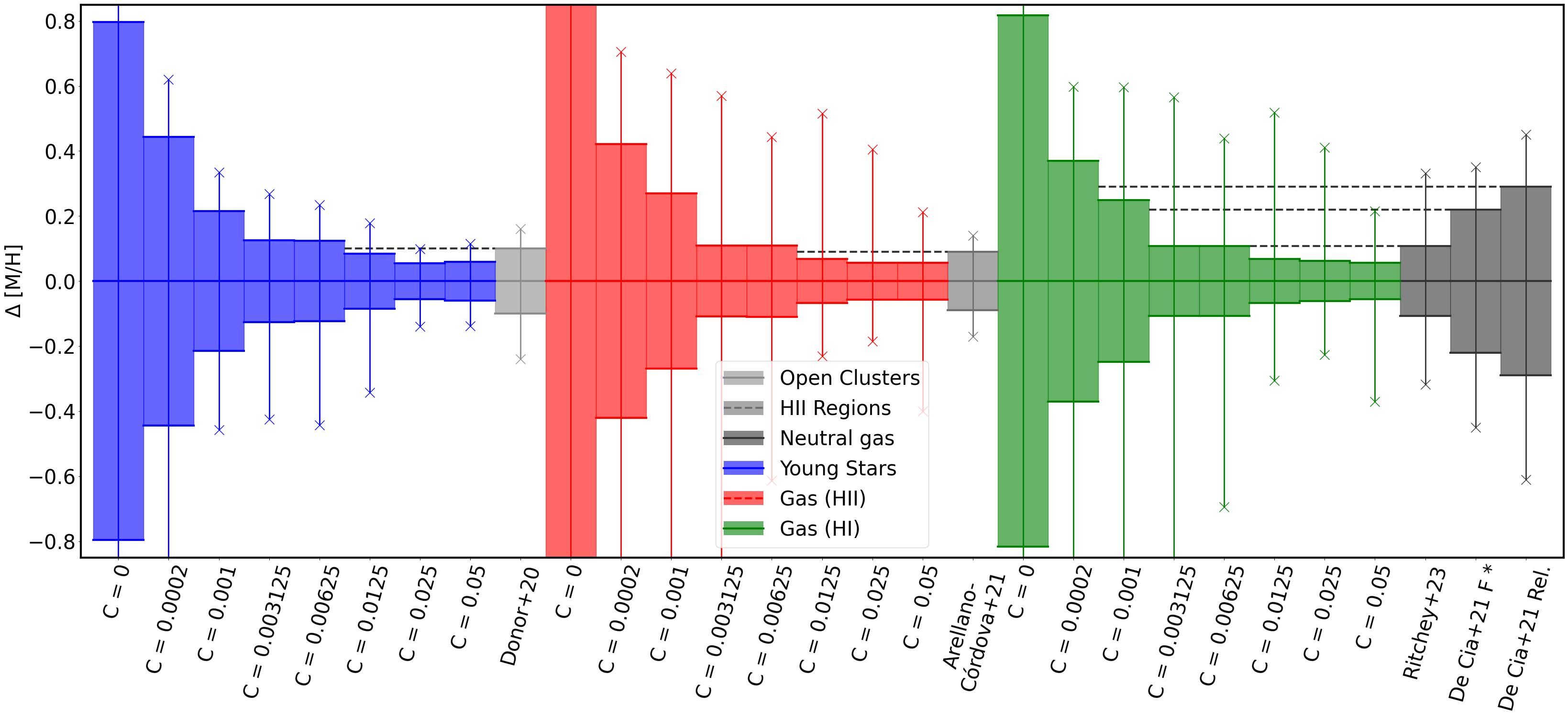}
    \caption{Comparison between the dispersion of [M/H] based on different tracers in the solar neighbourhood, both in observations and simulations. Vertical shadowed areas show the one sigma regions. Crosses united by a vertical line mark the extreme values. These values are plotted with respect to the mean of the sample. Blue, Red and Green boxes represent the distribution of young stars (<400 Myr), HII and HI, respectively, in simulations with different C values, indicated on the x-axes. Other boxes depict observations: light grey corresponds to open clusters (\protect\citealt{Donor_2020}), grey boxes represent observations of HII regions (\protect\citealt{ArellanoCordova20,ArellanoCordova}) and dark grey is used to show data from \protect\cite{decia} and \protect\cite{Ritchey}. Horizontal dashed lines serve as references to ease simulation-observation comparison.} 
    \label{fig:boxplots}
\end{figure*}

Regarding metallicity, \cite{Rennehan_2021} showed that the addition of any sub-grid metal diffusion model into SPH simulations results in more realistic MDFs, and that more refinement is needed in order to reproduce observations. More recent works use observations to constrain this free parameter. Using non-cosmological simulations, \cite{bario} suggested a lower limit for C at 0.01 based on the Ba distribution in dwarf galaxies. \cite{zinc} state that C$\sim$0.01 is the adequate value in order to reproduce the Zn distribution of metal-poor stars in non-cosmological simulations of Local Group dwarf galaxies. \cite{escala} studied the width of the MDF in Local Group dwarfs using fully cosmological hydrodynamical simulations, and they reproduced observations using C $\sim$ 0.003 (note that they refer to this coefficient as C$_{0}$ in their paper). \cite{escala}'s work also stresses that varying C by an order of magnitude only effects the results slightly, in particular showing that the width of the MDF in one particular simulation remains almost constant when increasing C to 0.03.

In the original set of NIHAO simulations the scaling constant C of the metal mixing model is set to 0.05, following the work of \cite{shen} whose constraints were made with respect to heat diffusion. We evolve the same initial conditions for galaxy g8.26e11 using 7 different values of metal diffusion, using values between 0 and 0.025, in addition to comparing with data from the fiducial 0.05. All runs maintain the fiducial heat diffusion rates adopted in the NIHAO simulations. 

\subsection{Simulated Galaxy Properties}
\label{sec:properties}

Table \ref{tab:galproperties} lists general properties of the original and new simulations, and it can be seen how changing the diffusion coefficient C affects galactic properties. There is no strong change in the general properties or the simulated galaxies, consistent with what was shown in \cite{hopkins18} for FIRE simulations and \cite{arora2022} who used NIHAO Local group simulations with and without a metal diffusion rate of 0.05. We confirm that the morphology of our simulated galaxies remains essentially constant, with all scale lengths and scale heights of the thin and thick disc components varying by at most 12\% with respect to the mean value across the various simulations. On the other hand, Table \ref{tab:galproperties} shows that there is a very slight trend for simulations with a higher C value to have less gas content and more stars. This is consistent with an interpretation of metal mixing making gas cooling more efficient, which would facilitate more star formation (see also \citealt{hopkins18}), although this trend is not reproduced in all simulations, which may be due to the stochasticity of star formation.

\subsection{Post Processing}
\label{sec:HI}
 To calculate HI fractions within each gas particle, we post process the simulations to include self-shielding, as outlined in Appendix 2 of \cite{rahmati13a}. We also include effects of radiation coming from star forming regions using equation \ref{eqrahmati} from \cite{rahmati13b}, which relates star formation to gas surface density (represented by hydrogen column density, $N_{\text{H}}$) and defines the rate of photoionisation of the HI gas ($\Gamma_{*}$), allowing a determination of HII, which we also use in this study. Note that the amount of HI and HII are somewhat sensitive to such modelling of ionisation rates, with neutral hydrogen masses around a factor of two larger using these models than when assuming that the ISM is optically thin.

 \begin{equation}
 \label{eqrahmati}
     \Gamma_{*} \sim 8.5\times 10^{-14}\text{s}^{-1} \left( \frac{N_{\text{H}}}{10^{21} \text{cm}^{-2}} \right)^{0.4}
 \end{equation}

  The galaxy haloes are identified using Amiga Halo Finder \citep[\textit{AHF},][]{ahf} with halo masses defined within a sphere of r=r$_{\mathrm{vir}}$.

\section{Data}
\subsection{Observations of stars and gas in the MW's disc}
\label{sect:obs}

\cite{Donor_2020} provide observed metallicities for 10 young (< 400 Myr) open star clusters with galactocentric radii between 5-11 kpc (we only use those flagged as high quality). The source of the metallicity values is the chemical abundance data of individual stars from APOGEE DR16 \citep{apogeedr16}.

\cite{ArellanoCordova20,ArellanoCordova} present a homogeneous sample of 42 HII regions covering the MW's disc from 4-17 kpc in galactocentric radii. They provide O/H calculations using line intensity measurements ([OII] $\lambda\lambda$3726, 3729 and [OIII] $\lambda\lambda$4959, 5007), with which they calculate the O/H radial gradient. We use a subset of 20 HII regions closer than 3 kpc from the Sun to study the dispersion of [M/H] in the solar neighbourhood, as done in \cite{Esteban_2022}.
 %Metallicities of HII regions are calculated by \cite{Esteban_2022} from O/H ratios by subtracting 12+$\log_{10}$(O/H)$_{\odot}$=8.69, the photospheric O/H ratio recommended by \cite{asplund09}.
 
 \cite{decia} provides 25 observations of neutral ISM gas, using absorption through lines of sight to targets within 3 kpc from the Sun. In particular, they measured absorption in lines associated with the elements Ti, Cr, Fe, Ni, and Zn. They provide two different metallicity values for each target corresponding to two dust depletion calculations: the F$^{\ast}$ method \citep{Jenkins_2009} and the relative method \citep{De_Cia_2016}.
 
\cite{Ritchey}'s similar study extends to 4 kpc from the Sun with 84 targets, of which only 7 were further than 3 kpc from the Sun. They use the F$^{\ast}$ method to correct for dust depletion. They propound that metallicities obtained in \cite{decia} are underestimated because they are derived from elements that are amongst the most refractory ones (with the exception of Zn). These elements are less reliable for the study of the metallicity of the ISM because they are more affected by dust depletion. For this reason, the metallicities calculated by \cite{Ritchey} are derived using only more volatile elements: B, C, N, O, Mg, Si, P, Cl, Mn, Cu, Zn, Ga, Ge, As, Kr, Cd, Sn, and Pb.

\subsection{Sample selection from MW simulations}
\label{sec:sample}
To compare with the aforementioned observations, we select different samples of gas and stars from our simulations, always spanning a range of $\pm$2 kpc from the plane of the disc. All simulated samples are selected to reproduce the galactocentric radial distributions of the observational datasets they are compared to.

For the comparison with \cite{Donor_2020}, we consider simulated young star particles with ages smaller than 400 Myr, covering a range of galactocentric radii from 5 kpc to 11 kpc.

To match the HII regions from \cite{ArellanoCordova20,ArellanoCordova}, we select gas particles within galactocentric radii from 4-17 kpc when calculating the O/H radial gradients, and between 5-11 kpc when studying the metallicity dispersion in the solar neighbourhood. Averages and standard deviations are weighted by the HII content of the gas particles, considering only the ionisation caused by stellar radiation (as explained in Section \ref{sec:HI}).

To replicate the \cite{decia} and \cite{Ritchey} observations, we use a sample of gas particles whose galactocentric radius ranges from 4-12 kpc, in this case weighting the particles by their neutral hydrogen content.

\begin{figure}
    \centering
    \includegraphics[width=\columnwidth]{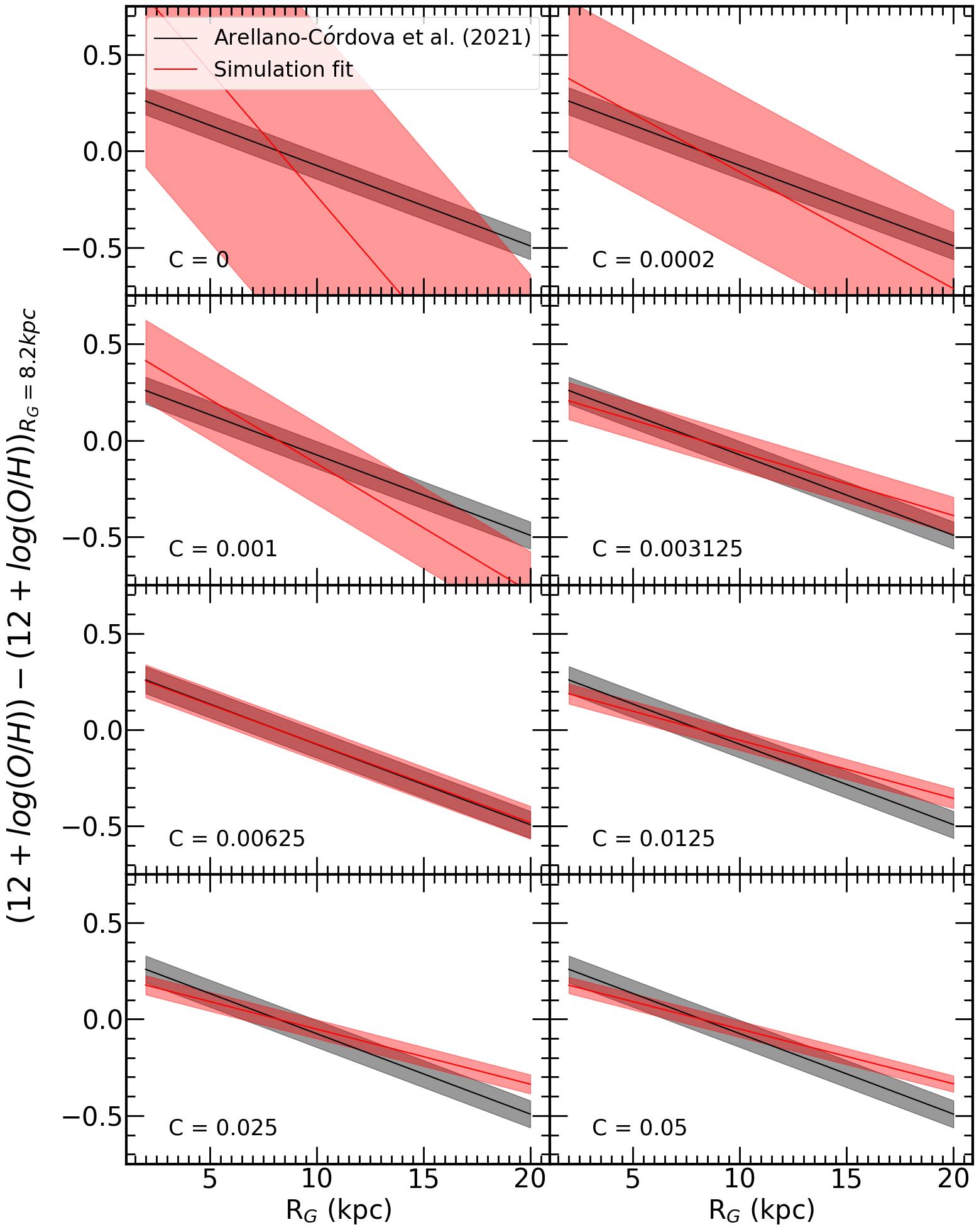}
    \protect\caption{Gradients of 12 + log(O/H)  vs. galactocentric radius, for HII regions in the disc of  simulated galaxies (red) and of the observed MW (black, from \citealt{ArellanoCordova}). Each individual panel contains the data of one simulation with a different diffusion coefficient C. The solid line is obtained through a linear fit of the data, and the shadowed area shows the dispersion, calculated as the mean of absolute differences between the data and the value predicted by the linear fit. Gradients are normalised with respect to the value of 12 + log(O/H) at R$_{G}$ = 8.2 kpc. Absolute values are found in table \protect\ref{tab:grads}.}
    \label{fig:gradients}
\end{figure}

\section{Results}
\label{sec:results}
Fig. \ref{fig:boxplots} presents a comparison of the dispersion of total metals, [M/H], for the simulations with various values of mixing rate C, indicated in the x-axes, as well as for different observational tracers in the solar neighbourhood. Metallicities are normalised to the mean values of the solar region in each case, in order to compare the metallicity dispersions rather than its absolute value. The one sigma dispersions are shown as the shaded regions, whilst crosses mark the extreme values in each sample. 

The left (blue) boxes represent the distribution of young stars (<400 Myr) in simulations with different C values, the middle (red) boxes show simulated HII gas, whilst the right (green) boxes show simulated HI gas. These simulated metallicities are compared to the corresponding observations of open clusters \citep[light grey,][]{Donor_2020}, HII regions \citep[grey,][]{ArellanoCordova20,ArellanoCordova} and HI gas \citep[dark grey][]{decia,Ritchey}. Dashed horizontal lines serve as references to facilitate simulation-observation comparison. Table \ref{tab:boxplots} shows the absolute data. Sizes of samples in the simulations are indicated by their total mass. Taking into account the typical mass of star particles in the simulations ($5\times10^{4}$ M$_{\odot}$), the masses of young star samples in the table correspond to $\sim$100-400 star particles depending on the simulation.

Looking at simulation results in Fig. \ref{fig:boxplots}, it can be seen that metal dispersion decreases rapidly as C increases, as result of mixing becoming more efficient. The standard deviation of the sample selected in simulations can be compared to the corresponding sets of observations (we do not compare maximum and minimum values since they are affected by the number of objects, which is relatively small in the observations). Overall, in the case of young stars and HII regions, as well as for neutral gas when comparing with \cite{Ritchey}'s data, values of C of 0.00315 and 0.0125, including the in-between value of 0.00625, provide a quite good match to observations. Instead, lower values of C give too much dispersion. In particular, the C=0 case shows clearly the need for implementing metal diffusion in Lagrangian codes. Oppositely, values of C greater than 0.0125 give too little dispersion to match the observations. 

Results of \cite{decia} for neutral gas stand out as having more dispersion than other observations. A low metal mixing rate with a value of C of 0.001, or even less, would provide sufficient mixing of metals to match these observations.

%\cite{Esteban_2022} studied the O/H number abundance ratio gradient with galactocentric radius in the MW using HII regions data from \cite{ArellanoCordova}. They fit observations linearly to find the slope of the gradient. We compare this gradient with those in our simulated galaxies. Aside from the gradient itself, another value to compare is the dispersion of the gradient, defined as the mean absolute difference of the value of the O/H number abundance ratio with the value obtained from the fit. In order to reproduce observations of HII regions we select gas in the disc of the simulated galaxies (-2 kpc < z < 2 kpc) and we weight both the linear fit and the dispersion of the relation by HII content.
%
%with low temperatures (T < 1000K) and high densities (n > 0.1 cm$^{-3}$). The election of the temperature and density cuts is performed by plotting all %simulated gas in the T-n space and selecting the bulk of dense cold gas that we interpret as candidate for resembling HII regions.
%

Fig. \ref{fig:gradients} shows the radial dependence of the Oxygen to Hydrogen (O/H) abundance ratio, comparing observations (grey) as compiled in \cite{Esteban_2022}, to those fitted in our simulations (red). The 1 sigma dispersions are shaded. Table \ref{tab:grads} presents the values of such gradients and dispersions. As expected from above, the dispersion of the relation is smaller for simulations with stronger diffusion (higher C). The simulation with C = 0.00625 is the one with a most similar dispersion of the relation to \cite{ArellanoCordova} observations, although again C=0.003125 and C=0.0125 also provide a good match to observations.

Furthermore, we can observe a general trend that simulations with stronger metal mixing have a smaller (less negative) O/H gradient. A value of C=0.00625 matches reasonably well the slope of the O/H gradient reported by \cite{ArellanoCordova}. %As C increases to values of C=0.0125 or higher, the radial slope O/H in simulations becomes lower than in the observations.

\begin{figure}
    \centering
    \includegraphics[width = \columnwidth]{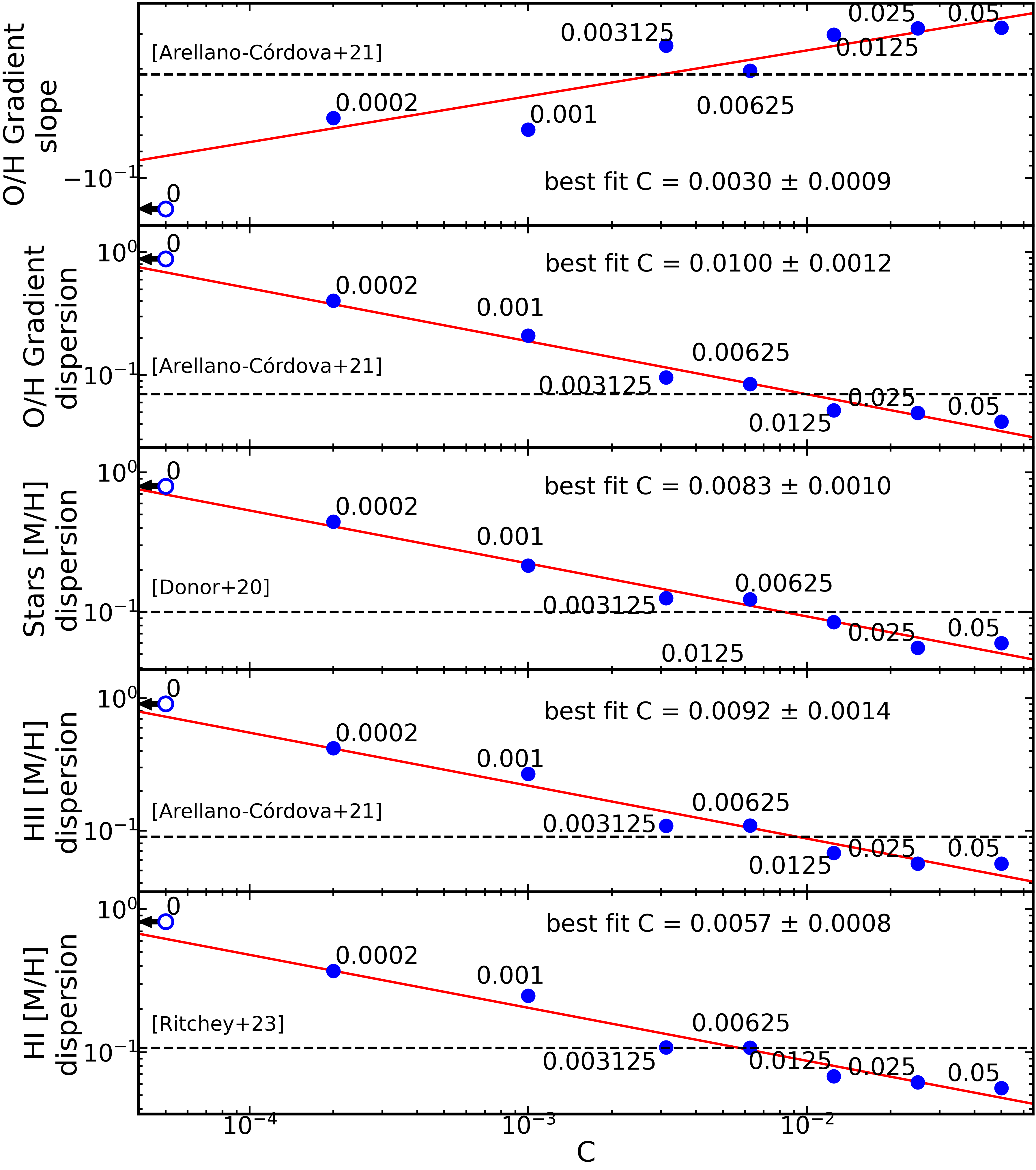}
    \caption{Power-law fits to different studied quantities in simulations with respect to the diffusion coefficient C. Red lines show the fits to the data marked by blue points. Each blue circle corresponds to a simulation with a C value written next to it. Empty circles show points at C = 0 which were not used for fitting. Horizontal dashed lines show the observed values in the MW together with the respective literature reference. Each panel displays the C value at which the fitted trend reproduces observations.}
    \label{fig:fits}
\end{figure}

Finally, in Fig. \ref{fig:fits} we fit each observable (A) with a power-law dependence on the diffusion coefficient C ($A = a \times C^{b}$), with fits (red lines) to the data (blue points). The observables used are: O/H gradient slope and dispersion from Table \ref{tab:grads}, young star, HII and HI [M/H] dispersion as reported in Table \ref{tab:boxplots} column $\sigma$ [M/H]. The dashed lines represent MW observed values. We can then find the value of C from the fit that best matches the observations, with uncertainties calculated through the fitted parameters. The derived values of the best fit C, across all observables, range from 0.003 to 0.01. A weighted average of the coefficients obtained for the five different observables through this method gives C = 0.0064 $\pm$ 0.0004. %Repeating this process but leaving the O/H gradient slope out of the calculation leaves C = 0.0077 $\pm$ 0.0005.

\section{Discussion \& conclusions}
\label{sec:disc}
In the hydrodynamical simulations of a MW analogue galaxy studied here, mixing is modelled using a trace-free shear tensor (equations \ref{newdiffrate}), with rates controlled by a diffusion coefficient C. A suite of 8 simulations with values of C ranging from 0 to 0.05 were run and analysed. It is clear from the results that metal diffusion is required, with the C=0 case resulting in too much inhomogeneity in the [M/H] values of several tracers, such as young stars, HII and HI gas, compared to observations (Fig. \ref{fig:boxplots}). Our results also show that the mixing rate adopted has a large effect on the dispersion of metals, with a general trend of smaller dispersion for larger mixing coefficient. These two results confirm what has been shown in multiple studies \citep[e.g.][]{shen,alfascattering,escala,Rennehan_2021}. 

Our results suggest that adopting the fiducial value of C from the NIHAO simulations, C=0.05, results in too much mixing, with an ISM that is too homogeneous (Fig. \ref{fig:boxplots}), and a radial metallicity gradient that is too shallow (Fig. \ref{fig:gradients}). This is consistent with findings in \cite{Buck_2021}, who found that adopting such a metal diffusion rate leads to a metallicity dispersion for stars in the [O/Fe] vs [Fe/H] plane, which is significantly lower than what is observed.

It is also apparent from our results that one can match any single observed metallicity dispersion by making an appropriate choice of mixing rate in the simulation. However, matching a single observation is never sufficient. What is required in galaxy formation simulations is to self-consistently match a range of observations. In the adopted model, our results suggest that a value of C in the region of around C=0.003-0.01 is able to match the observed (in)homogeneity of the MW ISM, except for the observations of \cite{decia}. 

This range of C values is similar to what was required (C=0.003-0.03) to match the width of the MDF of local dwarf galaxies \citep{escala}, using cosmological simulations of dwarf galaxies run with the FIRE \citep{hopkins14,hopkins18} model. They found that using a value of C that is ten times lower, C=0.0003, results in an MDF that is significantly too wide. This suggests that adopting a value of C that matches the inhomogeneities reported by \cite{decia} would not only result in a mismatch with other observed values of the ISM, but also with constraints provided by the width of the MDF of dwarf galaxies. 

Our results also show that a power law reproduces the dependence of metal dispersion in young stars, HII regions and neutral gas, as well as the gradient of O/H and its dispersion, with the diffusion coefficient C, as shown in Fig. \ref{fig:fits}. Using this power law fit, we are able to find a weighted average for the value of C that best matches each of the five observables, resulting in a value of 0.0064. 

The power law relation naturally explains why increasing C by a factor of ten was found to have relatively small effects on dispersion in FIRE simulations \citep{escala,hopkins18}. Our derived best value of C=0.006 is inside the range of values (C=0.003-0.03) for which \cite{escala} found agreement between FIRE simulations and observational data of dwarf galaxies in the Local Group.

Despite the aforementioned agreement between our results and \cite{escala}'s, we note that a very precise calibration of the mixing strength using observations would not necessarily yield the same result for different simulation codes. In the adopted model, mixing strength depends on the resolution of the simulation (see equation \ref{newdiffrate}), and different codes for hydrodynamical simulations use a variety of different galaxy formation models and hydrodynamics implementations which may produce different results from the same initial conditions. Thus, the same set of observables tracking the scatter of chemical abundances may be fit by simulations run with a different code and resolution using different metal mixing strengths, even if the same sub-grid metal diffusion model is being used. Finding a calibration process that covers such a large space of variables is extremely challenging, and we leave it for future work.

\begin{table}
\caption{Absolute data of observed and simulated samples in the MW. D20: \protect\cite{Donor_2020}, AC21: \protect\cite{ArellanoCordova}, DC21: \protect\cite{decia}, RI23: \protect\cite{Ritchey}.}
\label{tab:boxplots}
\resizebox{\columnwidth}{!}{%
\begin{tabular}{ccccccc}
\hline
Object type                                                                                         & Source    & \# of objects & Mean {[}M/H{]} & $\sigma$ {[}M/H{]} & {[}M/H{]}$_{min}$ & {[}M/H{]}$_{max}$ \\ \hline
\begin{tabular}[c]{@{}c@{}}Young\\ Clusters\end{tabular}                                     & D20       & 10           & -0.04          & 0.10               & -0.28             & 0.12                                                                                   \\ \hline
HII Regions                                                                                  & AC21      & 25           & -0.18          & 0.09               & -0.35             & -0.04                                                                             \\ \hline
\multirow{3}{*}{\begin{tabular}[c]{@{}c@{}}ISM\\ Neutral Gas\end{tabular}}                   & DC21 Rel. & 20           & -0.17          & 0.29               & -0.78             & 0.28                                                                                \\
                                                                                             & DC21 F*   & 20           & -0.24          & 0.22               & -0.69             & 0.11                                                                                           \\
                                                                                             & RI23      & 84            & 0.02          & 0.10               & -0.22             & 0.24                                                                                         \\ \hline
                                                                                              & Diffusion  & Mass ($10^{7}$ M$_{\odot}$)  & &  & &  \\ \hline
\multirow{8}{*}{\begin{tabular}[c]{@{}c@{}}Young\\ Stars\\ (\textless{}400 Myr)\end{tabular}} & C=0        & 0.66           & -0.18          & 0.80               & -2.40             & 0.89              \\
                                                                                              & C=0.0002   & 1.63          & 0.01           & 0.44               & -1.29             & 0.63              \\
                                                                                              & C=0.001    & 0.72           & 0.15           & 0.21               & -0.30             & 0.49              \\
                                                                                              & C=0.003125 & 2.88          & 0.17           & 0.13               & -0.26             & 0.43              \\
                                                                                              & C=0.00625  & 3.34          & 0.14           & 0.12               & -0.31             & 0.37              \\
                                                                                              & C=0.0125   & 3.25          & 0.18           & 0.08               & -0.17             & 0.35              \\
                                                                                              & C=0.025    & 2.27          & 0.21           & 0.06               & 0.07              & 0.31              \\
                                                                                              & C=0.05     & 1.17          & 0.24           & 0.06               & 0.11              & 0.36              \\ \hline

\multirow{8}{*}{HII Gas}                     & C=0        & 52.99        & -0.44          & 0.91               & -3.18             & 0.95              \\
                                                                                              & C=0.0002   & 31.17        & 0.04           & 0.42               & -1.86             & 0.74              \\
                                                                                              & C=0.001    & 47.18        & 0.07           & 0.27               & -1.40             & 0.71              \\
                                                                                              & C=0.003125 & 36.20        & 0.17           & 0.11               & -0.78             & 0.74              \\
                                                                                              & C=0.00625  & 43.61        & 0.16           & 0.11               & -0.45             & 0.60              \\
                                                                                              & C=0.0125   & 41.05        & 0.18           & 0.07               & -0.05             & 0.70              \\
                                                                                              & C=0.025    & 32.40        & 0.22           & 0.06               & 0.03              & 0.63              \\
                                                                                              & C=0.05     & 31.98       & 0.22           & 0.06               & -0.18             & 0.43              \\ \hline
\multirow{8}{*}{HI Gas}                                                                      & C=0        & 348.38        & -0.06          & 0.82               & -3.27             & 0.94              \\
                                                                                              & C=0.0002   & 281.57        & 0.15           & 0.37               & -1.30             & 0.74              \\
                                                                                              & C=0.001    & 416.88        & 0.13           & 0.25               & -1.30             & 0.73              \\
                                                                                              & C=0.003125 & 296.88
        & 0.17           & 0.11               & -0.86             & 0.74              \\
                                                                                              & C=0.00625  & 379.84
        & 0.16           & 0.11               & -0.53             & 0.60              \\
                                                                                              & C=0.0125   & 353.87
        & 0.18           & 0.07               & -0.12             & 0.70              \\
                                                                                              & C=0.025    & 287.54
        & 0.21           & 0.06               & -0.01             & 0.63              \\
                                                                                              & C=0.05     & 314.96
        & 0.22           & 0.06               & -0.15             & 0.43              \\  \hline

\end{tabular}%
}

\end{table}

% \usepackage{graphicx}
\begin{table}
\caption{Parameters of the linear fits to 12 + log(O/H) data against galactocentric radius for calculating the radial gradient. This table contains the fitted slopes, the values of the fitted relations at the solar radius, and the dispersions of the fits.}
\label{tab:grads}
\resizebox{\columnwidth}{!}{%
\begin{tabular}{cccccccc}
\hline
\multicolumn{8}{c}{Simulations}                                                                                                                                                                                                            \\ \hline
C        & Slope  & \begin{tabular}[c]{@{}c@{}}12 + log(O/H)\\ at R$_{G}$=8.2kp\end{tabular} & \multicolumn{1}{c|}{Dispersion}  & C       & Slope  & \begin{tabular}[c]{@{}c@{}}12 + log(O/H)\\ at R$_{G}$=8.2kp\end{tabular} & Dispersion \\ \hline
0        & -0.129 & 8.21                                                                     & \multicolumn{1}{c|}{0.88}        & 0.00625 & -0.041 & 9.00                                                                     & 0.08       \\
0.0002   & -0.060 & 8.79                                                                     & \multicolumn{1}{c|}{0.44}        & 0.0125  & -0.030 & 9.03                                                                     & 0.05       \\
0.001    & -0.066 & 8.90                                                                     & \multicolumn{1}{c|}{0.21}        & 0.025   & -0.029 & 9.05                                                                     & 0.05       \\
0.003125 & -0.033 & 9.00                                                                     & \multicolumn{1}{c|}{0.10}        & 0.05    & -0.028 & 9.06                                                                     & 0.04       \\ \hline
\multicolumn{8}{c}{Observations (Arellano-Córdova+21)}                                                                                                                                                                                     \\ \hline
\multicolumn{3}{c}{Slope}                                                                    & \multicolumn{3}{c}{12 + log(O/H) at R$_{G}$=8.2kpc} & \multicolumn{2}{c}{Dispersion}                                                        \\ \hline
\multicolumn{3}{c}{-0.042 $\pm$ 0.009}                                                       & \multicolumn{3}{c}{8.50}                            & \multicolumn{2}{c}{0.07}                                                              \\ \hline
\end{tabular}%
}

\end{table}

The main results of this work are listed here:

\begin{itemize}
    \item metal mixing (C>0) is essential for reproducing observations (Fig. \ref{fig:boxplots}), in line with conclusions drawn by \cite{Rennehan_2021};
    
    \item simulations with C = 0.003125, 0.00625 and 0.0125 yield good results when compared to observations of metal dispersion in young star clusters, HII regions and neutral gas (Fig. \ref{fig:boxplots}), as well as with the radial gradient and dispersion of O/H in HII regions (Fig. \ref{fig:gradients}). The results are consistent with all observations except for \cite{decia}, who found significant variability in the [M/H] values of the neutral ISM surrounding the Sun;

    \item we have established that the relationship between metal dispersion in young stars, HII regions and neutral gas, as well as the gradient of O/H and its dispersion, versus the diffusion coefficient C, can be described by a power law (Fig. \ref{fig:fits}). 
    
     \item by fitting a power law to the aforementioned quantities we have identified a range of C values, 0.003 to 0.01, that best reproduce observations, and have calculated a weighted average which provides a value of C = 0.0064 $\pm$ 0.0004. 
\end{itemize}

Such a value of C is one order of magnitude smaller than the one used in several present hydrodynamical SPH simulations. 
A well constrained metal mixing rate not only has implications for modelling the ISM, as shown in this study. Metal mixing rates also affect the metalicity of the circum-galactic medium, as well as the stellar MDF and by implication the number of low metallicity stars, allowing more robust comparison with populations in the various components of the Milky Way, in local dwarf galaxies, and with the metallicity distribution of stars and gas in external galaxies. 

\section*{Acknowledgements}
JSA thanks the Spanish Ministry of Economy and Competitiveness (MINECO) for support through a grant P/301404 from the Severo Ochoa project CEX2019-000920-S. CB is supported by the Spanish Ministry of Science and Innovation (MICIU/FEDER) through research grant PID2021-122603NBC22. ADC is supported by a Junior Leader fellowship from ‘La Caixa’ Foundation (ID 100010434), code LCF/BQ/PR20/11770010. We thank A.V. Macciò for sharing with us the run and initial conditions for g8.26e11. This research made use of computing time available on HPC systems at the Instituto de Astrofisica de Canarias. The authors thankfully acknowledge the technical expertise and assistance provided by the Spanish Supercomputing Network (Red Española de Supercomputación), as well as the computer resources used: the LaPalma Supercomputer, located at the Instituto de Astrofisica de Canarias. The freely available software \textit{pynbody} \citep{pynbody} has been used for part of this analysis.
\section*{Data Availability}
All data and simulations used in this work are available upon request. 
%%%%%%%%%%%%%%%%%%%%%%%%%%%%%%%%%%%%%%%%%%%%%%%%%%



\bibliographystyle{mnras}
\bibliography{example} % if your bibtex file is called example.bib

%\begin{thebibliography}{99}
%\end{thebibliography}

%%%%%%%%%%%%%%%%%%%%%%%%%%%%%%%%%%%%%%%%%%%%%%%%%%

%%%%%%%%%%%%%%%%% APPENDICES %%%%%%%%%%%%%%%%%%%%%

%%%%%%%%%%%%%%%%%%%%%%%%%%%%%%%%%%%%%%%%%%%%%%%%%%

% Don't change these lines
\bsp	% typesetting comment
\label{lastpage}
\end{document}